\documentclass[aps,prb,showpacs,reprint,unsortedaddress]{revtex4-1}

\usepackage{hyperref}
\usepackage{array}
\usepackage{dcolumn}
\usepackage{graphicx}
\usepackage{color}
\usepackage{amsmath}
\usepackage{amssymb}
\usepackage{amsfonts}
\usepackage{epstopdf}
\usepackage{dsfont}
\DeclareMathOperator{\Tr}{Tr}

\begin{document}

\title{Temperature dependence of the NMR spin-lattice relaxation rate for spin-1/2 chains}

\author{E. Coira}
\affiliation{DQMP, University of Geneva, 24 Quai Ernest Ansermet, 1211 Geneva,  Switzerland}
\author{P. Barmettler}
\affiliation{Supercomputing Systems AG, Technoparkstrasse 1, 8005 Z\"urich, Switzerland}
\author{T. Giamarchi}
\affiliation{DQMP, University of Geneva, 24 Quai Ernest Ansermet, 1211 Geneva,  Switzerland}
\author{C. Kollath}
\affiliation{HISKP, University of Bonn, Nussallee 14-16, 53115 Bonn, Germany}

\begin{abstract}
We use recent developments in the framework of time dependent matrix product state method (tMPS) to compute the nuclear magnetic resonance (NMR) relaxation rate $1/T_1$ for spin-1/2 chains under magnetic field and for different Hamiltonians (XXX, XXZ, isotropically dimerized). We compute numerically the temperature dependence of the $1/T_1$. We consider both gapped and gapless phases, and also the proximity of quantum critical points. At temperatures much lower than the typical exchange energy scale our results are in excellent agreement with analytical results, such as the ones derived from the Tomonaga-Luttinger liquid (TLL) theory and bosonization which are valid in this regime. We also cover the regime for which the temperature $T$ is comparable to the exchange coupling. In this case analytical theories are not appropriate but this regime is relevant for various new compounds with exchange couplings in the range of tens of Kelvin. For the gapped phases, either the fully polarized phase for spins chains or the low magnetic field phase for the dimerized systems, we find an exponential decrease in $\Delta/(k_BT)$  of the relaxation time and can compute the gap $\Delta$. Close to the quantum critical point our results are in good agreement with the scaling behavior based on the existence of free excitations.
\end{abstract}

\maketitle

\section{Introduction} \label{sec:introduction}
Quantum spin systems can exhibit a very rich set of phases, although they are usually described by very simple Hamiltonians\cite{auerbach_book_spins}. These range from phases with long range magnetic order to spin liquids, or even phases for which the order is more complex and rests on non-local order parameters. Understanding the behavior of such systems is thus an extremely challenging task with potential use for quantum computation or quantum simulation of other types of systems\cite{ward_simulator_TLL_review}. Among all spin systems, low dimensional ones such as spin chains and ladders are particularly interesting since quantum effects are large. In one dimension interactions between the excitations lead to various exotic states, ranging from gapped phases to phases possessing quasi-long range magnetic order known as Tomonaga-Luttinger liquids\cite{giamarchi_book_1d} (TLL).

In order to examine the various types of order that can be present, it is important to have a set of probes sensitive to correlation functions of the system. Fortunately a set of such probes such as neutron scattering, electronic spin resonance, Raman scattering and Nuclear Magnetic Resonance (NMR)\cite{abragam_rmn,slichter_rmn} exists. The NMR allows for various measurements such as the Knight shift, measuring the local magnetic field, or the so-called $T_1$ relaxation time, essentially sensitive to the decay of the local spin-spin correlation functions.

Although the principle of what is measured by the ratio $1/T_1$ is simple and relates to local spin-spin correlations, the theoretical determination is far from trivial. Very often various schemes of approximations of the exact formula are used. The first approximation consists in assuming that the NMR frequency is low enough (usually in the hundred of MHz range) compared to the temperature and that it can safely be set to zero\cite{abragam_rmn,slichter_rmn}. The second approximation is usually to reduce the local correlation function, which is a sum over all momenta, to a sum taken around special momentum values (e.g.~the $q\sim0$ values and values around either the antiferromagnetic wavevector $q\sim\pi/a$, where $a$ is the lattice spacing, or a corresponding incommensurate one when the magnetization is finite). This last approximation is reasonable when the characteristic scale of excitations (typically the temperature) is low enough compared to the magnetic exchange, so that the excitations around these wavevectors can be well separated. Finally, to compute the correlations, some continuous approximations such as bosonization, exploiting the above points, are usually employed. This set of approximations has allowed a connection between NMR measurements and theoretical predictions for quantum chains and ladder systems.

In the recent years a successful set of magnetic systems, in which the magnetic exchanges are considerably lower\cite{giamarchi_nature_2008,zapf_becreview} than in previously used materials, typically around 10K, has been developed. These materials have the advantage over previously studied ones that they can be manipulated by the application of experimentally reachable magnetic fields, from zero magnetization to full saturation. This tunability opens the possibility to investigate new physics such as the universality of the TLL\cite{klanjsek_bpcp,dora_nmr_nanot,sato_nmr}, and to use these magnetic systems as quantum simulators\cite{ward_simulator_TLL_review} for other quantum systems such as itinerant bosons\cite{zapf_bec_dtn,bouillot_dynamical_ladder,mukhopadhyay_BPCB,schmidiger_dimpy_spectra,jeong_dimpy_nmr}. Since in these materials the exchange energy scale is now much closer to the typical measurement temperatures, it invalidates partly, or pushes to very low temperatures, the above mentioned approximations. Thus, in connection with this new class of materials, a direct method to compute the NMR relaxation time without having to resort to these approximations is needed.

This is what we undertake in the present paper, by using a time dependent matrix product state method (tMPS) \cite{Vidal2004,WhiteFeiguin2004,DaleyVidal2004,Schollwoeck2011} to compute directly the relaxation time at finite temperature \cite{White_finT,Verstraete_finiteT_DMRG,zwolakVidal2004,barthel_tdmrg_finiteT}. An alternative approach would be the direct calculation of the correlation functions in the frequency domain as for example in Ref.~\onlinecite{tiegel_freqdyn}.

However, here we have chosen to use the time-dependent method to evaluate the time dependence of the local spin-spin correlations, since these can be directly related to the ratio $1/T_1$. We calculate the quantity as a function of the temperature, even in regimes for which the temperature is not negligible compared to the magnetic exchange. Previous works have shown numerical results for autocorrelations and relaxation time obtained via DMRG techniques\cite{naef_diffusion,sirker_NMR_tDMRG} or exact diagonalization methods\cite{fabricius_diffusion}. 

The plan of the paper is as follows. In Section~\ref{sec:models} we briefly introduce the models we consider by discussing their Hamiltonians and phase diagrams. Section~\ref{sec:oneovertone} defines and discusses the spin-lattice relaxation mechanism, its relation to spin-spin correlation functions, and some analytical results valid in the low temperature limit which we will use to benchmark our numerical results. In Section~\ref{sec:procedure} we describe the procedure adopted for the numerical computation. We then move in Section~\ref{sec:results} to the results for the different models considered, namely XXZ spin chains and dimerized ones. We show how the results of the numerical calculations connect with the standard field theoretical approaches for $1/T_1$. Section~\ref{sec:conclusion} presents conclusions and perspectives. In the Appendices we give additional details about the computations and some preliminary tests made to check the robustness of the code.

\section{Models} \label{sec:models}

We consider spin-1/2 chains characterized by different anisotropies of the coupling, or even dimerization. In addition, a magnetic field is applied along the $z$-direction. The first model we investigate is the antiferromagnetic XXZ chain, whose Hamiltonian is given by
\begin{equation} \label{eq:XXZham}
 H=J\sum_j\left[\frac{1}{2}\left(S^+_j S^-_{j+1}+\text{h.c.}\right) + \Delta S^z_jS^z_{j+1}\right]-h\sum_j S^z_j,
\end{equation}
where $S^\alpha_j=\frac{1}{2}\sigma^{\alpha}_j$ is a spin operator for a spin $1/2$ on site $j$, $\alpha=x,y,z$ denotes its direction, and $\sigma^\alpha$ the Pauli matrices. $S^\pm_j= S^x_j\pm i S^y_j$ are the spin rising and lowering operators. The parameter $J$ gives the spin coupling strength, $\Delta$ is dimensionless and measures the anisotropy, $h$ is the amplitude of the applied magnetic field along the $z$ direction. The $g$ factor, the Bohr magneton and $\hbar$ have been absorbed into $h$ and $J$, which both have here the dimensions of an energy.
\begin{figure}
  \centering
  \includegraphics[width=0.45\textwidth]{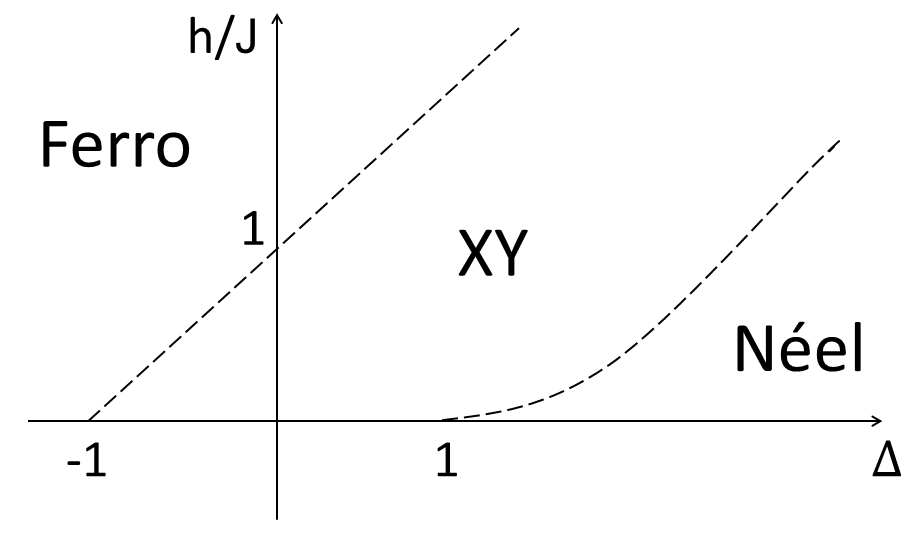}
  \caption{\label{fig:phasedi} Schematic phase diagram at zero temperature of the XXZ model (Eq.~\ref{eq:XXZham}) as a function of the magnetic field $h$ and of the anisotropy parameter $\Delta$. 'Ferro' stands for the phase in which the spins are ferromagnetically aligned along the $z$ direction. $XY$ denotes a massless phase with dominant in-plane antiferromagnetic correlations which is a TLL\cite{giamarchi_book_1d}. 'N\'eel' denotes an Ising antiferromagnetically ordered phase along $z$. The behavior of the boundary as a function of $h$ around the point $\Delta=1$ reflects the Berezinski-Kosterlitz-Thouless (BKT) behavior of the gap at the transition. After Fig.~1.5 in Ref.~\onlinecite{mikeskaKolezhuk2004}.}
\end{figure}
For the isotropic case $\Delta=1$, the model corresponds to the Heisenberg (or XXX) Hamiltonian, while for $\Delta=0$ we have the XX model which can be mapped via a Jordan-Wigner transformation \cite{giamarchi_book_1d} onto a free-fermion model with a fixed chemical potential. The phase diagram of the XXZ model is given in Fig.~\ref{fig:phasedi}. The boundary between the XY and ferromagnetic phases is given by $h_c=J(1+\Delta)$. The boundary between the XY and N\'eel phases is given by the triplet gap, which is a function of $\Delta$ \cite{mikeskaKolezhuk2004}. In this work we will limit ourselves to the case $0\leq\Delta\leq1$.
\begin{figure}
  \centering
  \includegraphics[width=0.45\textwidth]{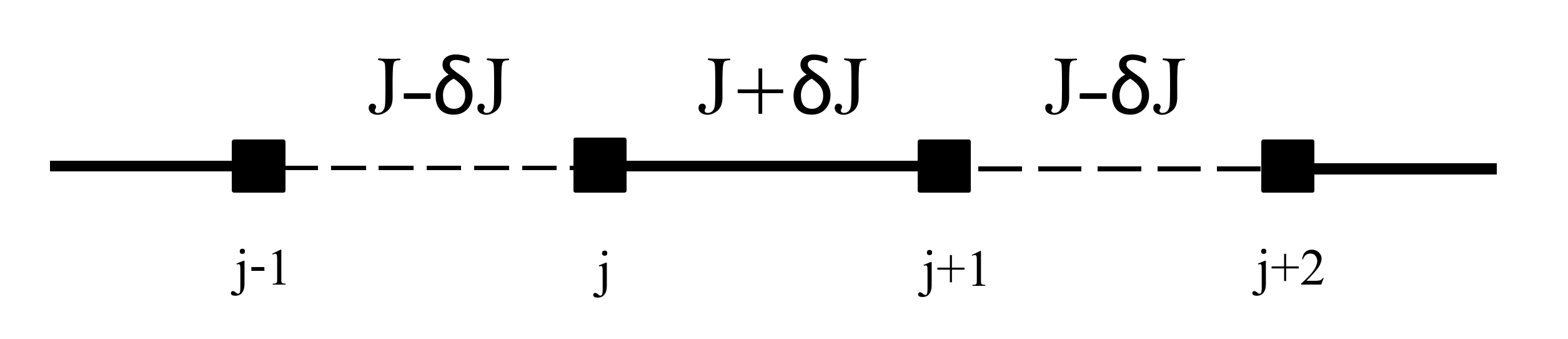}
  \caption{\label{fig:lattice_dim} Pictorial representation of the dimerized chain: a spin 1/2 is located on each black square, the strength of the coupling (between nearest neighbor spins only) is alternated with values $J_s=J+\delta J$ and $J_w=J-\delta J$.}
\end{figure}

\begin{figure}
  \centering
  \includegraphics[width=0.45\textwidth]{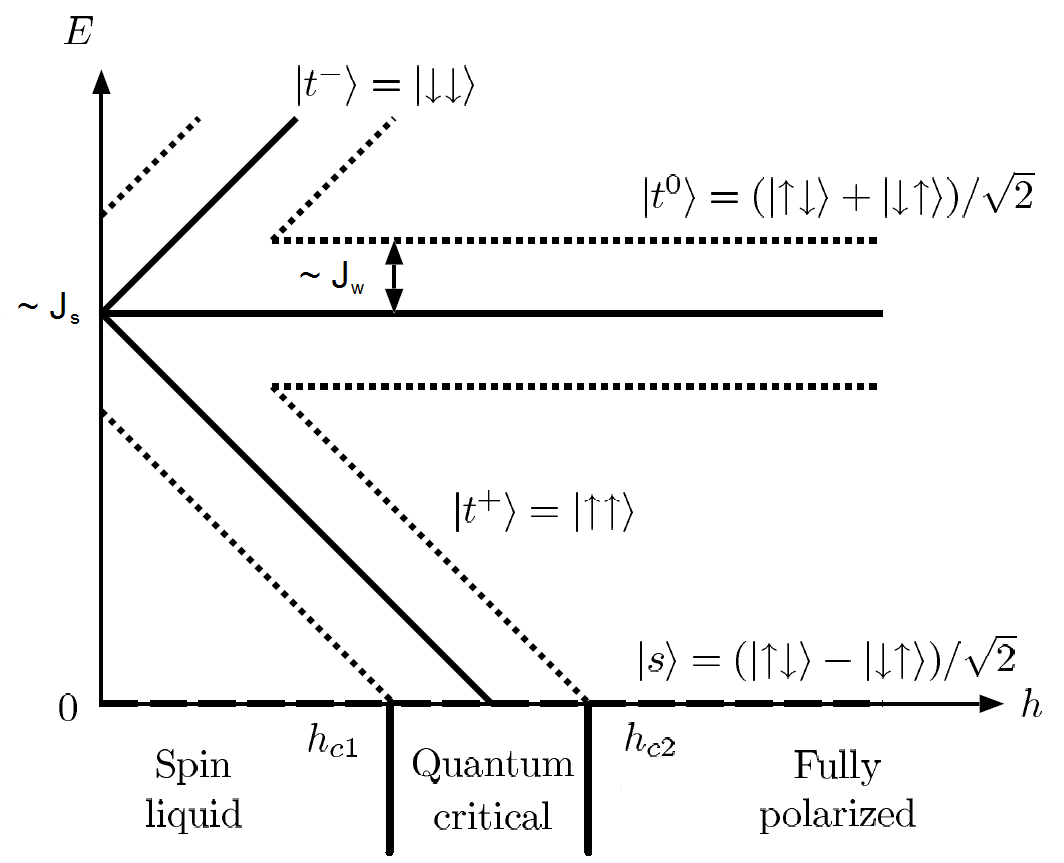}
  \caption{\label{fig:phasedi_dim} Sketch of the energy spectrum of excitations for the dimerized chain under the application of a magnetic field $h$ in the limit of large dimerization ($J\approx\delta J$). In this limit, the nature of excitations can be approximated by considering the state of two spins 1/2 on a strong bond. At $h=0$ there is an energy gap of the order of $J_s$  between the singlet and the three triplet states. The magnetic field $h$ splits the triplets and brings down the excitation energy of the state $\left|t^+\right\rangle$. Due to the presence of the weak bonds the triplets can be delocalized and, thus, have a dispersion in energy of the order of $J_w$ (the boundaries of which are represented by the dotted lines). At sufficiently high magnetic field, the energy of the lowest triplon band is close to the energy of the singlet state. This leads in the extended system to a quantum critical phase for $h> h_{c1}$ with gapless excitations. This phase exists up to the point $h_{c2}$ for which the triplon band is totally filled and a fully polarized phase arises. Picture adapted from Fig.~2 in Ref.~\onlinecite{bouillot_dynamical_ladder}}.
\end{figure}

Additionally, we consider the dimerized Heisenberg chain which is described by the Hamiltonian
\begin{equation} \label{eq:dimham}
 H=\sum_j\left(J+(-1)^j\delta J\right) \mathbf{S}_j\cdot \mathbf{S}_{j+1}-h\sum_j S^z_j.
\end{equation}
where $\mathbf{S}_j=(S^x_j,S^y_j,S^z_j)^T$ denotes the vector of the spin at site $j$. Here $J_s=J+\delta J$ is the strong exchange coupling on every second bond and $J_w=J-\delta J$ the weak coupling of the other bonds. A pictorial representation of this system is given in Fig.~\ref{fig:lattice_dim}. At zero magnetic field ($h=0$) such a model has a non-trivial spin-0 ground state (spin liquid) with a gap to the first excitation which is a band of spin-1 excitations (triplons)\cite{giamarchi_book_1d}. This is particularly easy to see in the limit of large values of the dimerization. In this limit, strongly, antiferromagnetically coupled spin dimers are formed due to the strong exchange coupling  $J_s$ on every second bond. These dimers are themselves coupled by the weaker interaction $J_w$. The lowest excitations are the one-triplon excitations which can be accurately described as a single dimer excited from spin 0 ($\left|s^{}\right\rangle$) to spin 1 ($\left|t^\pm\right\rangle$, $\left|t^0\right\rangle$), delocalized on the chain (see Fig.~\ref{fig:phasedi_dim}).
The role of the magnetic field along $z$ is to progressively close the gap to the excitations. At $h>h_{c1}$ a quantum critical phase arises with gapless excitations. For stronger magnetic fields the spins of the chain become polarized, and above the second critical magnetic field $h_{c2}$ the ground state is fully polarized with a gapped spectrum. For more details on dimerized chains see e.g.~Ref.~\onlinecite{giamarchi_book_1d} and references therein.

\section{Spin-lattice relaxation time $\bold{T_1}$} \label{sec:oneovertone}

We introduce in this section the so called {\it spin-lattice relaxation time} $T_1$, one of the important time-scales of NMR  measurements. In NMR experiments the nuclear spins of the sample, previously polarized by an applied magnetic field, are perturbed using an electromagnetic pulse. The time constant $T_1$ characterizes the process by which the component of the nuclear magnetization along the direction of the applied magnetic field (denoted here by $z$) reaches thermodynamic equilibrium with its surroundings (the lattice) after the perturbation\cite{abragam_rmn,slichter_rmn}. The evolution of the nuclear magnetization along $z$ is:
\begin{equation} \label{eq:origin}
M_z(t)=M_{z,eq} \left(1-e^{-t/T_1}\right).
\end{equation}
Quite generally in a solid the ratio $1/T_1$ can be related directly to the spin-spin correlations of the electronic system. Using the fact that the nuclear-electronic coupling (which is the hyperfine one) is very weak, one obtains\cite{slichter_rmn} the relation:
\begin{equation}
\frac{1}{T_1}=\frac{\gamma_n^2}2 \left[A_{\perp}^2 \left(S^{xx}(\omega_0)+S^{yy}(\omega_0)\right) + A_{\parallel}^2 S^{zz}(\omega_0)\right],
\label{eq:redfi}
\end{equation}
where $\gamma_n$ is the nuclear gyromagnetic ratio of the measured nuclear spin, $A_{\perp}$ and $A_{\parallel}$ are the longitudinal and transverse components of the hyperfine tensor, $S^{\alpha\alpha}(\omega_0)$ with $\alpha=x,y,z$
are the local spin-spin correlation functions at the nuclear Larmor frequency $\omega_0$, and
\begin{equation}
 S^{\alpha\alpha}(\omega_0) = \int_{-\infty}^{+\infty} dt\; e^{i\omega_0 t} \left\langle S^\alpha(x=0,t) S^\alpha(x=0,t=0)\right\rangle,
\label{eq:salphaalpha}
\end{equation}
where $\langle\rangle$ denotes the thermal and quantum average given by
\begin{equation}
 \langle\cdots\rangle=\frac{\Tr[e^{-\beta H}\cdots]}{\Tr[e^{-\beta H}]}.
\end{equation}
Note that in formula (\ref{eq:redfi}) we have implicitly assumed that the hyperfine coupling term was essentially $q$ independent. This covers a large number of cases, for example the ones in which the relaxation is measured on the site carrying the electronic spin. There are also interesting cases for which the $q$ dependence of the hyperfine term can filter some modes, for example the modes at $q=\pi$ if the relaxation is measured mid-point between two neighboring sites. This leads to different formulas and interesting properties\cite{sirker_hypermomdep,sirker_hypermomdep2,karrasch_hypermomdep}. Note that techniques similar to the ones used here but computing the finite temperature, space and time dependent spin correlations allow to treat this problem as well. We leave this more complicated case for further studies, and focus here to the generic case for which the local spin-spin correlation is sufficient.   

The first two terms in Eq.~(\ref{eq:redfi}) can be conveniently expressed in terms of the $S^+$ and $S^-$ operators
\begin{equation}
 S^{xx}(\omega_0)+S^{yy}(\omega_0) = \frac12 \left[S^{+-}(\omega_0)+S^{-+}(\omega_0)\right]
\end{equation}
The time integral over infinite time is only valid theoretically, since neither in the experiment nor in the simulation one could expect doing the sum over an infinite interval of time. In practice, two time scales compete. One is the typical time $t \sim 1/\omega_0$ above which one can expect the oscillations coming from the frequency $\omega_0$ to become strong and regularize the integral. The second time scale hidden in the correlation itself is the decay of the correlation linked to the finite temperature. Since typical NMR frequencies are of the order $\omega_0 \simeq 20~\text{MHz}$ while the typical lowest temperatures at which such experiments are done are of the order of $40~\text{mK}\simeq 790~\text{MHz}$, for all practical purposes we can expect that the decay due to the temperature regularizes the integral. We will thus in the following give the expression by taking this usual limit $\omega_0 \to 0$ and keeping in Eq.~\ref{eq:salphaalpha} a finite integration domain up to a maximum time $t_0$. We will see that this time is important not only from the numerical point of view, but also because in some cases, at high enough temperatures, the approximation of setting the frequency $\omega_0$ to zero leads to divergences.

Using the approximations discussed above, one obtains that
\begin{equation} \label{eq:nmrfinal}
\begin{split}
S^{\lambda\mu}(\omega_0\rightarrow0) &\simeq \int_{-t_0}^{+t_0} dt\; \left\langle S_j^\lambda(t)S_j^\mu(0)\right\rangle \\
               &=  2 \int_0^{+t_0} dt\; \operatorname{Re}\left\langle S_j^\lambda(t)S_j^\mu(0)\right\rangle.
\end{split}
\end{equation}
Now $(\lambda,\mu)$ can be $(\pm,\mp)$ or $(z,z)$. Since we have set $\omega_0 = 0$ inside the integral in the above expression, and considered that $e^{\beta\omega_0}\simeq1$ as explained above, the two time integrals of $+-$ and $-+$ correlations also become identical (note of course that this is not the case for the correlations themselves at finite time). We can thus compute the one that is the most convenient numerically depending on the specific case.

Since in this work we focus on the parameter dependence of generic Hamiltonians we will omit the factors $\gamma_n^2A_{\perp}^2$ and $\gamma_n^2A_{\parallel}^2$, which depend on the specific material. For a specific material they have to be considered and in general both terms might be important. However, in this work we are not focusing on a specific material and have chosen to consider for each example only one of the terms separately.
We thus compute numerically
\begin{align}
\left(\frac{1}{T_1}\right)_{\pm\mp} &= 2 \int_0^{+t_0} dt\; \operatorname{Re}\left\langle S_j^{\pm}(t)S_j^{\mp}(0)\right\rangle, \label{eq:nmrpm}\\
\left(\frac{1}{T_1}\right)_{zz} &= 2 \int_0^{+t_0} dt\; \operatorname{Re}\left[\left\langle S_j^z(t)S_j^z(0)-m^2\right\rangle\right]. \label{eq:nmrzz}
\end{align}
Note that with the definitions in Eq.~\ref{eq:nmrpm} and Eq.~\ref{eq:nmrzz} the units of $1/T_1$ become time and not one over time as for the original definition in Eq.~\ref{eq:origin}.

The NMR relaxation rate $1/T_1$ in one dimension can be computed in the low energy TLL representation\cite{giamarchi_book_1d}. This calculation is valid when the temperature is low enough compared to the typical spin energy scales. In that case, neglecting the subdominant temperature corrections and the $zz$ contribution in Eq.~\ref{eq:redfi} (small compared to the $+-$ term), one finds \cite{bouillot_dynamical_ladder}
\begin{multline}\label{eq:oneovert1_analy}
\frac{1}{T_1} \ \simeq \ \lim_{\omega_0 \to 0}-\frac{2}{\beta\omega_0}\operatorname{Im}\chi_{+-}^R(\text{x=0},\omega_0) \ \simeq\\ \simeq \ \frac{4A_x\cos\left(\frac{\pi}{4K}\right)}{u}\left(\frac{2\pi k_B T}{u}\right)^{\frac{1}{2K}-1}B\left(\frac{1}{4K},1-\frac{1}{2K}\right),
\end{multline}
where $u$ and $K$ are the TLL parameters associated to the model, and $A_x$ is the amplitude coefficient relating the microscopic spin operator ${\bf{S}}_i$ on the lattice with the operators in the continuous field theory.  These coefficients have been computed both analytically and numerically in various contexts ranging from chains to ladders\cite{lukyanov_sinegordon_corr,giamarchi_coupled_ladders,hikihara_amplitude_ladder,hikihara_amplitude_chainh,bouillot_dynamical_ladder}. $\chi_{+-}^R(\text{x=0},\omega_0)$ is the retarded, onsite, $S^{+-}$ correlation function at the frequency $\omega_0$ (for the $q$ resolved susceptibility see Refs.~\onlinecite{cross_spinpeierls,schulz_correlations_1d,giamarchi_book_1d}). Note also that in this formula $\hbar$ and the lattice spacing has been set to one, thus omitted.

Eq.~\ref{eq:oneovert1_analy} has provided a quantitative estimation of the NMR in ladder systems for which the relaxation time could be measured\cite{klanjsek_bpcp,jeong16_TLLlad}. It will thus provide both a benchmark for the numerical evaluation of the relaxation time, as well as an estimation of the deviation from these ideal low energy properties.

\section{Numerical procedure} \label{sec:procedure}

As discussed in the previous section we need to compute correlation functions of the form
\begin{equation}
\left\langle \hat{B}(t)\hat{A}(0)\right\rangle_T = \Tr\left(\hat{\rho}_{\beta}\hat{B}(t)\hat{A}\right).
\label{eq:dmrg_corr1}
\end{equation}
Here $\hat{B}$ and $\hat{A}$ are spin operators with the relation $\hat{A}=\hat{B}^{\dagger}$. The expectation values of the operators are taken with the finite temperature density matrix
\begin{equation}
 \hat{\rho}_{\beta} = e^{-\beta H}/Z_{\beta},
\end{equation}
where $Z_\beta= \Tr (e^{-\beta H})$ and the inverse temperature $\beta=1/(k_BT)$. The time evolution of the operators is represented in the Heisenberg picture with $\hat{B}(t)=e^{iHt/\hbar}\hat{B}e^{-iHt/\hbar}$.

In order to use the matrix product state (MPS) representation at finite temperature\cite{White_finT,Verstraete_finiteT_DMRG,zwolakVidal2004,barthel_tdmrg_finiteT,Schollwoeck2011}, the density matrix $\hat{\rho}_{\beta}$ is encoded by a corresponding \emph{purification}\cite{nielsenchuang_book_qcomp} which is a pure state
in an enlarged Hilbert space:
\begin{equation}
 \hat{\rho}_{\beta}\longrightarrow\left|\rho_{\beta}\right\rangle\in\mathcal{H}\otimes\mathcal{H}_{\text aux}~,
\end{equation}
such that
\begin{equation}
 \Tr_{\text aux}\left|\rho_{\beta}\right\rangle\left\langle\rho_{\beta}\right|=\hat{\rho}_{\beta}.
\end{equation}
We choose the auxiliary space $ \mathcal{H}_{\text aux}= \mathcal{H}$ and $\Tr_{\text aux}$ denotes the trace over this space.
An MPS representation of $\left|\rho_{\beta}\right\rangle$ can be obtained by applying an imaginary time evolution, starting from the maximally entangled state
\begin{equation}
 \left|\rho_0\right\rangle\propto\sum_{\mathbf{\sigma}}\left|\mathbf{\sigma}\right\rangle\otimes\left|\mathbf{\bar{\sigma}}\right\rangle_{\text aux}
\end{equation}
with $\bar{\sigma}$ denoting the state not equal to $\sigma$. This maximally entangled state corresponds to the physical infinite temperature state $\hat{\rho}_0\propto\mathds{1}$, i.e. if one traces out the auxiliary degrees of freedom one obtains the identiy. Further, in each term the state $\left|\mathbf{\bar{\sigma}}\right\rangle_{\text aux}$ is chosen such that the magnetization is conserved in the following calculations, which enlightens considerably the numerical effort which needs to be spent.

\begin{figure}
  \centering
  \includegraphics[width=0.42\textwidth]{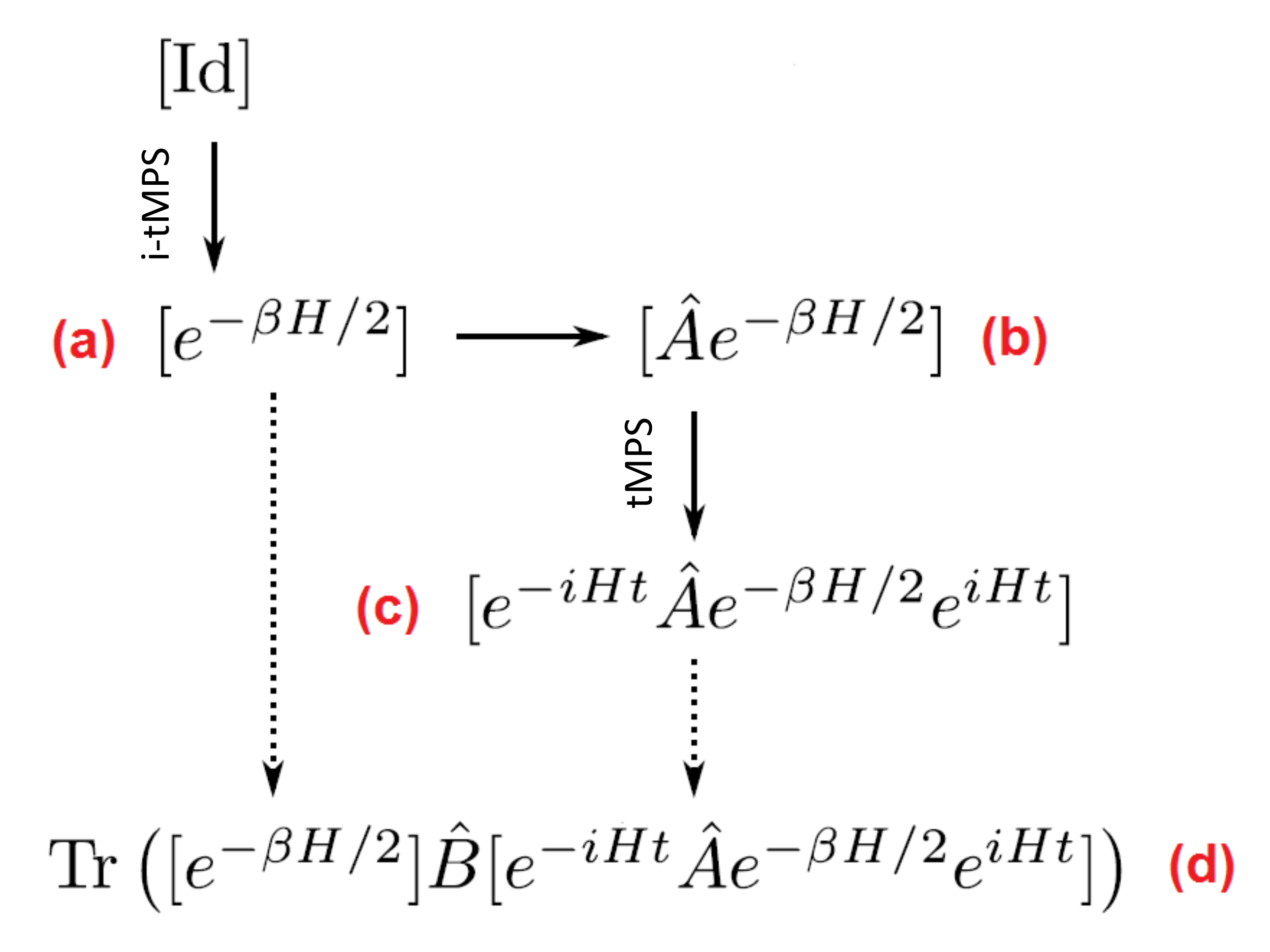}
  \caption{Representation of the scheme used for the computation of dynamical correlations at finite temperature. The initial state at finite temperature is prepared via imaginary time evolution \textbf{(a)}. A copy is created. The operator $\hat{A}$ is applied on this copy \textbf{(b)} and then a double real time evolution \textbf{(c)} is performed. At each time step the second operator is measured by sandwiching it through the two resulting states \textbf{(d)}. This gives us the desired correlation. After Fig.~2 in Ref.~\onlinecite{barthel_tdmrg_finiteT} with $\hbar=1$.}
  \label{fig:schemeB}
\end{figure}

Using the cyclic property of the trace, and expliciting the time dependence of the operator $\hat{B}$ and the density matrix,
one can rewrite Eq.~\ref{eq:dmrg_corr1} as
\begin{equation}
 \left\langle\hat{B}(t)\hat{A}\right\rangle_T=\frac{1}{Z_{\beta}}\Tr\left([e^{-\beta H/2}]\hat{B}[e^{-iHt/\hbar}\hat{A}e^{-\beta H/2}e^{iHt/\hbar}]\right).
\label{eq:dmrg_corr2}
\end{equation}
The square brackets indicate which parts of this expression are approximated as an MPS \cite{barthel_tdmrg_finiteT}. The bracketing is not unique and several different approaches exist. However, we found this one tested in Ref.~\onlinecite{karrasch_tdmrg_finiteT} to be often the most efficient for the here considered correlations. The approximation of the bracketed operators is calculated using an imaginary and real time evolution and the application of the local operators $\hat{A}$ and $\hat{B}$. In Fig.~\ref{fig:schemeB} the scheme is sketched. In each step of the real or imaginary time evolution, the evolved operators are approximated by an MPS with bond dimensions as small as possible for a given constraint on the truncated weight. The convergence of our results with the chosen truncated weight is assured. Typical values for maximum bond dimension used here are up to $1000$ states. Depending on the magnitude of the singular values after each decomposition, we keep those which are bigger than a minimal truncation $\epsilon$. This has been chosen of the order of $10^{-20}$ for imaginary time evolution and $10^{-10}$ for real time evolution. 

As discussed in the previous section (see Eq.~\ref{eq:nmrfinal}), we are especially interested in onsite, dynamical spin-spin correlation functions at finite temperature T:
\begin{equation}
 \left\langle \hat{B}(t)\hat{A}(0)\right\rangle_T \longrightarrow\left\langle S^{\lambda}_j(t)S^{\mu}_j(0)\right\rangle_T,
\label{eq:dmrg_corr3}
\end{equation}
where $j$ is the site index and $(\lambda,\mu)=(\pm,\mp)$ or $(z,z)$. From now on for practical reasons we will denote
\begin{equation}
 \left\langle S_j^{\lambda}(t)S_j^{\mu}(0)\right\rangle_T = S^{\lambda\mu}_T(t).
\end{equation}
In order to access the desired results (Eqs.~\ref{eq:nmrpm} and \ref{eq:nmrzz}) the time integral of the real part of these correlations from $t=0$ to $t=+t_0$ is required. Numerical results are taken at discrete times which are multiples of the time-step $\delta t$ chosen for the real time evolution within tMPS. Typical values of the steps are $\delta\beta=0.01~J^{-1}$ and $\delta t=0.05~\hbar/J$ respectively for the imaginary and real time evolution, for the XXZ system. For the dimerized system we choose typically $\delta\beta\approx 0.01474~J_w^{-1}$ and $\delta t\approx 0.0737~\hbar/J_w$. The convergence with the time step of the time evolution is assured. The amplitude of the time step for real time evolution is chosen to be small enough to guarantee good approximation of the proper integral:
\begin{multline}
 2 \int_{0}^{+t_{\text{max}}} dt\; \operatorname{Re}S^{\lambda\mu}_T(t) \approx \\
\approx \sum_{l=1}^{N}\delta t\left[\operatorname{Re}S^{\lambda\mu}_T((l-1)\delta t)+\operatorname{Re}S^{\lambda\mu}_j(l\delta t)\right],
\label{eq:intter}
\end{multline}
where $N$ is the total number of time steps at which the correlations are evaluated, $\delta t$ is the amplitude of a single time step and $t_{\text{max}}=N\delta t$. Depending on the available computational resources and on the constraints on the desired precision, runs are stopped after a certain $t_{\text{max}}$. In many cases this $t_{\text{max}}$ is large enough such that correlations are practically zero for larger times. In other cases it is not possible due to the numerical complexity to reach such a large  $t_{\text{max}}$. In order to have an idea of the value of the extended integral, we extrapolate its value for $t_{max}\to +\infty$ and associate to it an error bar. In Appendix~\ref{app:errors} the details of the extrapolation method and the determination of the error bars are given.

The numerical results shown in this work are obtained for the XXZ model for a chain of size $L=100$ and the correlations are evaluated at the central site $j=50$. For the dimerized model $L=130$ and $j=65$. The system sizes were chosen such that the perturbations do not yet reach the boundary of the system for times up to $t_{\text{max}}$. The resulting finite system size effects are small compared to the uncertainties introduced by the finite cut off of the time-integral and are therefore neglected.

To test the accuracy of the described procedure, some calculations have been performed for the XX model and compared with exact analytical results in Appendix~\ref{app:XX}.

\section{Results} \label{sec:results}

In the following subsections we present our numerical results for the Heisenberg model, the XXZ model and the dimerized model.

\subsection{Heisenberg model}
\begin{figure}
  \centering
  \includegraphics[width=0.4\textwidth]{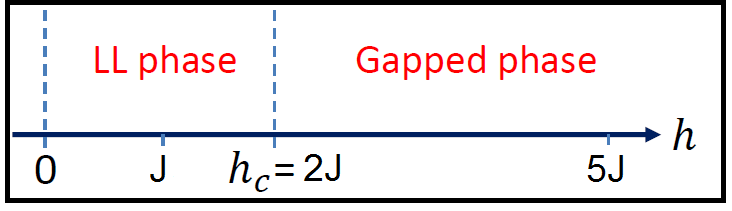}
  \caption{Phase diagram for the Heisenberg model ($\Delta=1$) as a function of the magnetic field $h$. For low magnetic field, the ground state is a TLL and the phase is gapless. In contrast, above the critical magnetic field $h_c=2J$ a ferromagnetic phase, which exhibits a gap in its excitation spectrum, arises.}
  \label{fig:XXX_phase}
\end{figure}

\begin{figure}
  \centering
  \includegraphics[width=0.51\textwidth]{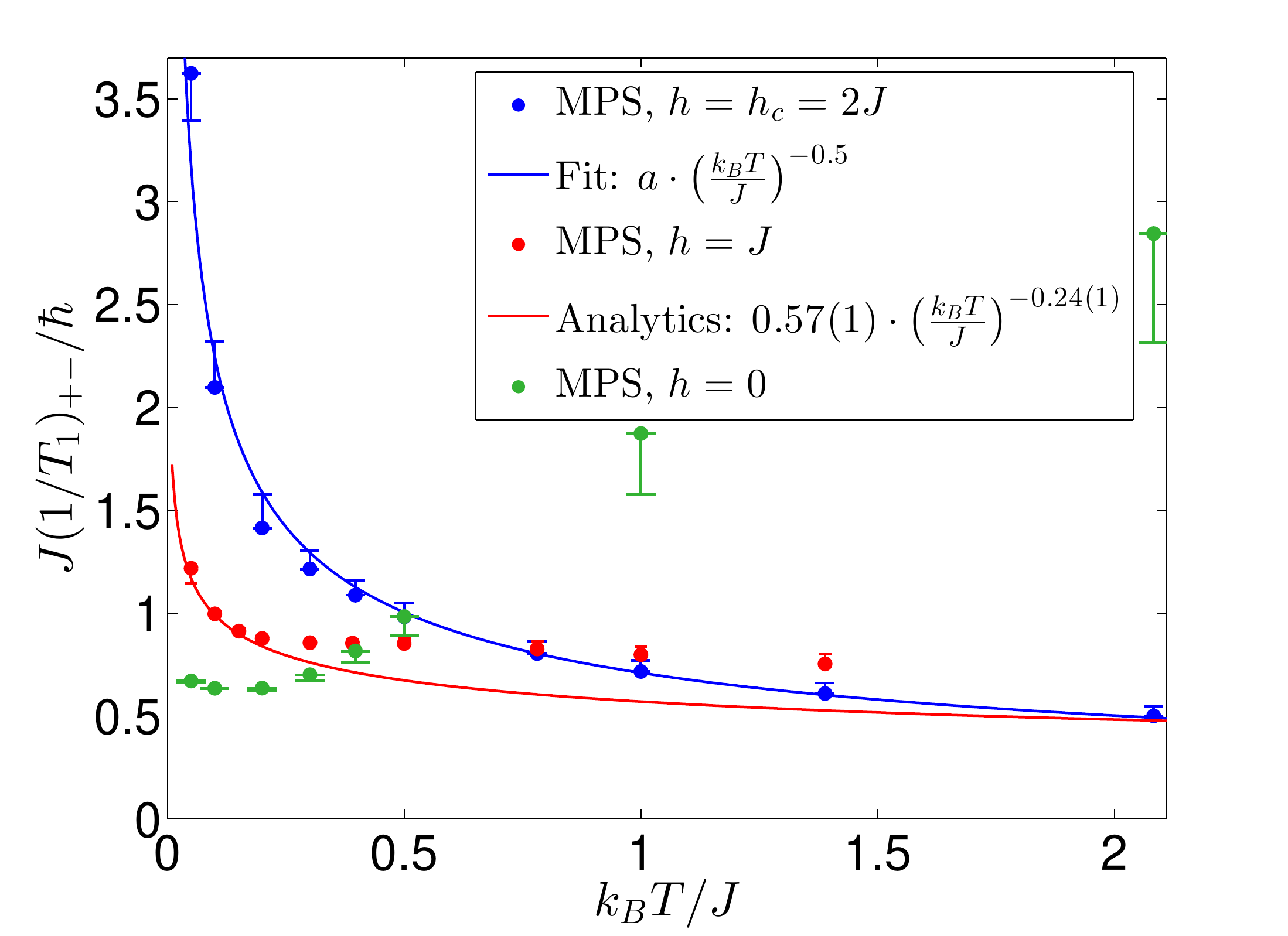}
  \caption{$(1/T_1)_{+-}$ as defined in Eq.~\ref{eq:nmrpm}, multiplied by $J/\hbar$ to have a dimensionless quantity, as a function of $k_BT/J$ for an XXX model under magnetic field. Dots with error bars are MPS results, solid lines are the analytical predictions and/or fits. $a=0.71(2)$ (fit parameter).}
  \label{fig:XXXtutto}
\end{figure}

Let us start by considering the Heisenberg model, i.e.~the XXZ model in Eq.~\ref{eq:XXZham}, with isotropic coupling $\Delta=1$, of spins 1/2 under a magnetic field, applied along the $z$ direction. A pictorial representation of the phase diagram as a function of $h$ is given in Fig.~\ref{fig:XXX_phase}. At low magnetic field the ground state of the system is a gapless TLL, whereas above the critical magnetic field ($h_c=2J$) a gapped phase develops. In order to explore the different phases and the quantum critical point we focus in the following on the fields $h=0$ and $h=J$ in the gapless phase, $h=h_c=2J$ at the quantum critical point and $h=5J$ in the gapped phase.
In Fig.~\ref{fig:XXXtutto} the results for the $1/T_1$ relaxation rate of the $S^{+-}$ correlations are shown at the magnetic field values $h=0$, $h=J$ in the gapless phase, and $h=2J$ at the quantum critical point. For the numerical calculations on the Heisenberg model and the XXZ model we have chosen a chain of size $L=100$, a minimal truncation of $\varepsilon_{\beta}=10^{-20}$, a retained states maximum of 400, and steps $\delta\beta=0.01~J^{-1}$ for the imaginary time evolution. For the real time evolution we have chosen the minimal truncation $\varepsilon_t=10^{-10}$, a retained states maximum of 800, a maximal truncated weight of $10^{-6}$, a time step $\delta t=0.05\hbar/J$ and $t_{\text{max}}=30\hbar/J$.

The behavior of the relaxation time at zero magnetic field has been investigated previously both by analytical~\cite{schulz_spins,sachdev_nmr} and numerical methods such as the quantum Monte-Carlo using the maximum entropy method in order to continue to the real time axis\cite{sandvik_nmr_spinchain, starykh97_logs} and tDMRG methods \cite{sirker_NMR_tDMRG}. In the asymptotic low-temperature limit one obtains\cite{cross_spinpeierls,schulz_correlations_1d} $1/T_1=const$ if logarithmic corrections $\ln^{1/2} (1/(k_BT))$ are neglected. Since the behavior at $h=0$ has been studied in detail in \onlinecite{sandvik_nmr_spinchain,sirker_NMR_tDMRG}, we here show the $h=0$ case for $(1/T_1)_{+-}$ mainly for comparison. At temperatures above $k_BT/J\gtrsim0.5$  an almost linear increase of the relaxation time with increasing temperature can be seen. At low temperature $(1/T_1)_{+-}$ shows an almost constant behavior. At temperatures below $k_BT/J\lesssim0.2$ it even increases again while lowering the temperature. This behavior is consistent with the logarithmic corrections and has been analyzed in \onlinecite{sandvik_nmr_spinchain}. The rise at larger temperatures has been treated in \onlinecite{sirker_NMR_tDMRG} and has been found compatible with an exponential increase with a scale of the order of the magnetic exchange $J$.

In the TLL region ($h=J$), as described in Sec.~\ref{sec:oneovertone}, the low temperature behavior of the relaxation rate should be approximately described by an algebraic decay with the exponent $\frac{1}{2K}-1$, which is fully determined by the TLL parameter $K$\cite{giamarchi_coupled_ladders, klanjsek_bpcp}. At larger temperature a breakdown of this low energy prediction is expected. In Fig.~\ref{fig:XXXtutto} our numerical results for $(1/T_1)_{+-}$ are quantitatively compared to the TLL predictions. Up to temperatures of about $k_BT/J\approx0.2 $ the numerical points agree within the error bars with the prediction made in Eq.~\ref{eq:oneovert1_analy}. This comparison is achieved using previously extracted values for the Tomonaga-Luttinger parameter $K=0.66(1)$, the amplitude $A_x=0.119(1)$ from Refs.~\onlinecite{hikihara_amplitude_ladder,hikihara_amplitude_chainh}, and $u=1.298(5)~J/\hbar$ (lattice spacing equal to 1) extracted from separate calculations which we performed using standard finite-size DMRG methods. More details about the determination are given in Appendix~\ref{app:LLpara}. Thus, all parameters in Eq.~\ref{eq:oneovert1_analy} are fully determined. For temperatures larger than $k_BT/J\approx 0.2$ the numerical results are much higher than the decaying TLL prediction. This is to be expected and clearer in the $q$ space for which the local correlation can be seen as a sum over all $q$ points. The TLL formula only contains the part coming from one of the low energy $q$ points ($q=\pi$ in the absence of magnetic field). At higher temperatures other $q$ points start to contribute significantly to the sum. The numerical results even seem to show a slight maximum around $k_BT/J\approx 0.5$ and then remain more or less constant in value up to the shown maximal temperature.

\begin{figure}
  \centering
  \includegraphics[width=0.51\textwidth]{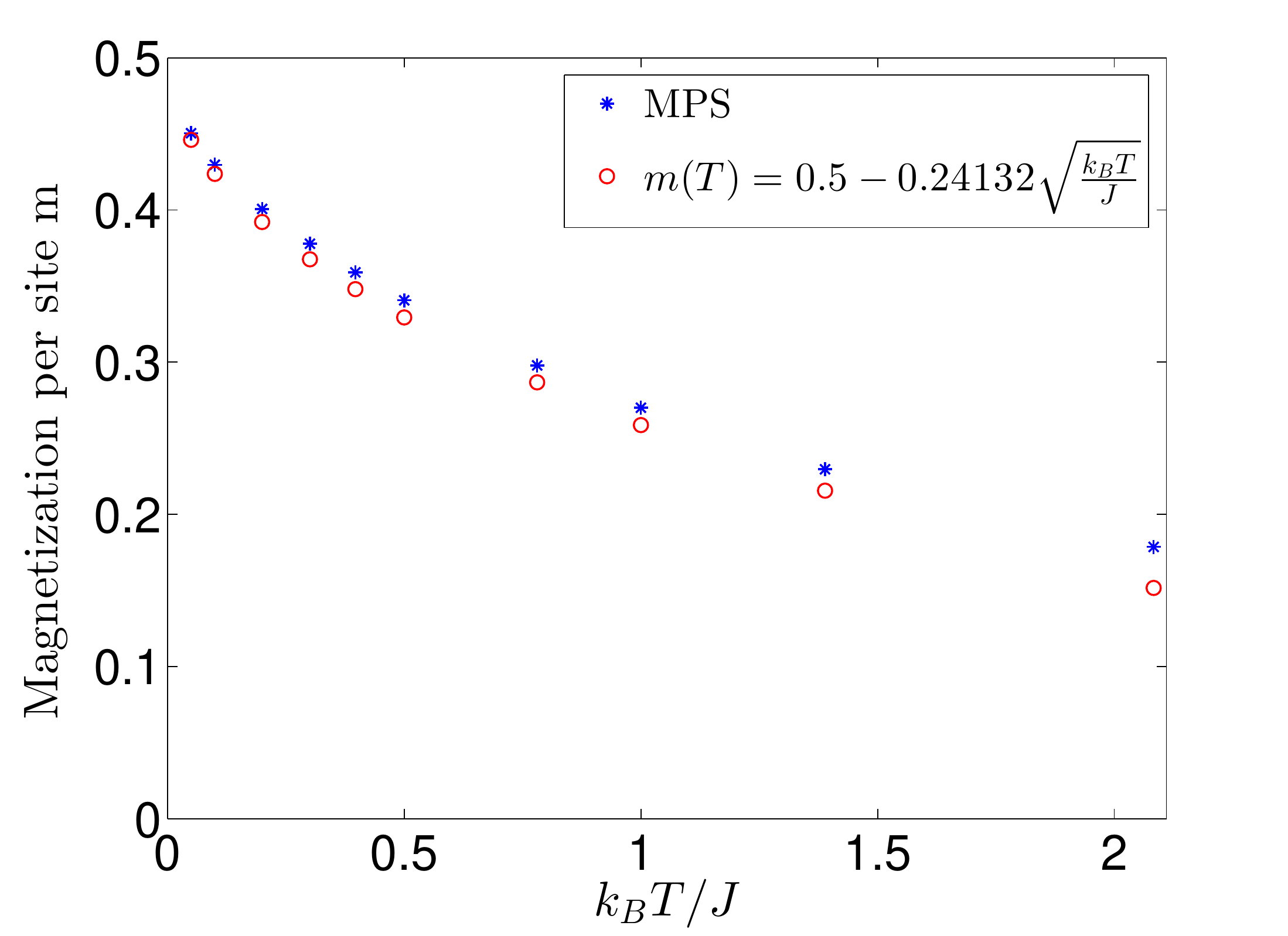}
  \caption{Magnetization per site as a function of $k_BT/J$ for the Heisenberg model at the critical field $h=2J$. Blue stars are MPS results (error bars are too small to be seen), red circles are the analytical predictions.}
  \label{fig:critmag}
\end{figure}

\begin{figure}
  \centering
  \includegraphics[width=0.51\textwidth]{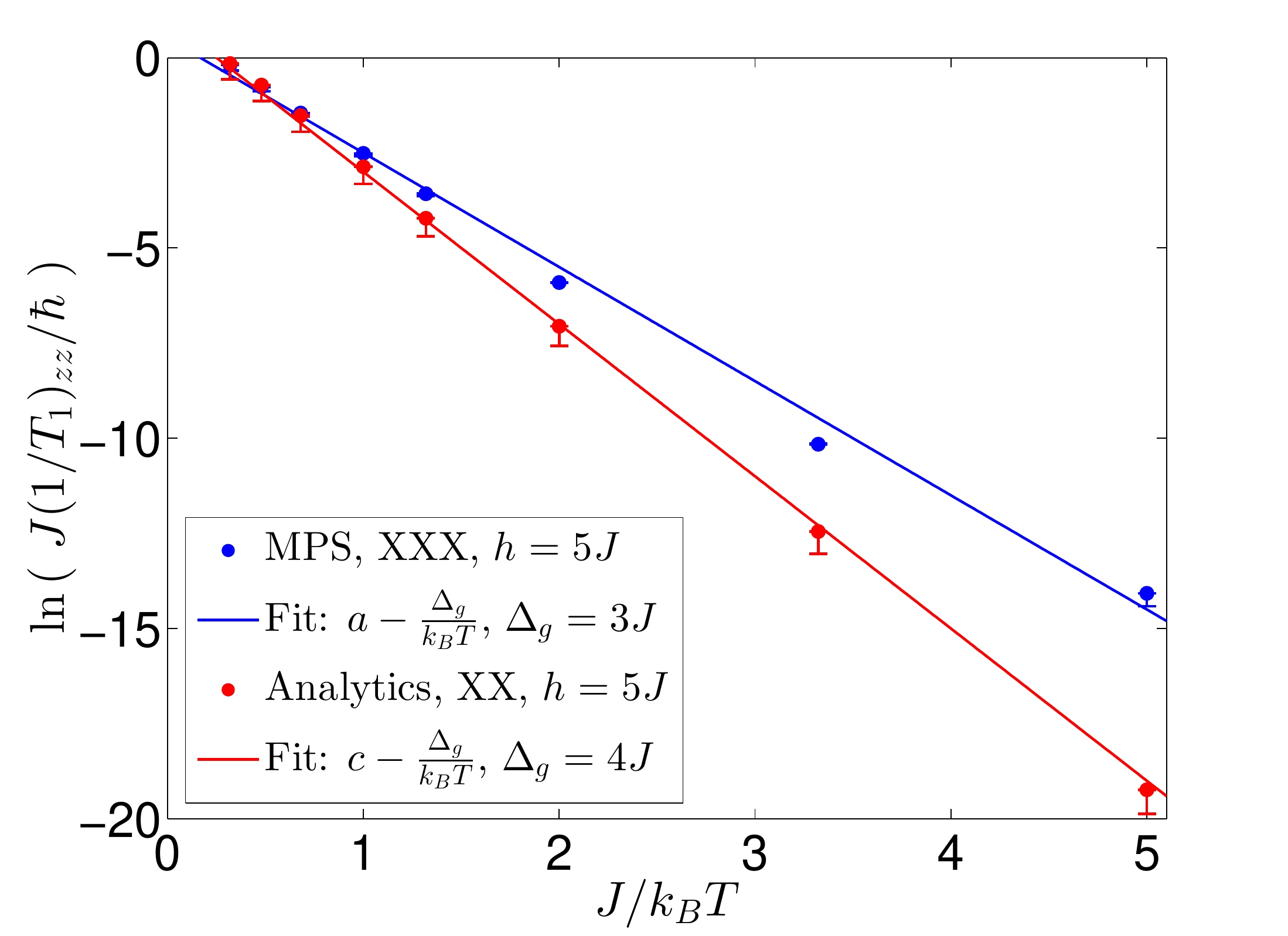}
  \caption{Logarithm of $J(1/T_1)_{zz}/\hbar$ (unitless quantity) from Eq.~\ref{eq:nmrzz} as a function of $J/k_BT$. We considered the Heisenberg and the XX model, with $h=5J$. Blue dots with error bars are MPS results, red dots with error bars are analytical results and solid lines are the fits according to the expected behaviors. $a=0.5(1)$ and $b=1.0(1)$ are the fit parameters. \label{fig:gappati}}
\end{figure}

At the quantum critical point  $h=2J$, an algebraic divergence of the $1/T_1$ with the temperature is also expected in the low T limit. It is predicted to behave as $\propto (k_BT/J)^{-0.5}$ as obtained  in Refs.~\onlinecite{Sachdev_qaf_magfield,Chitra_spinchains_field,orignac07_bec}. This behavior has been experimentally observed for example in the Heisenberg chain compound copper pyrazine in Ref.~\onlinecite{kuhne_coppyr_nmr}, and discussed in Ref.~\onlinecite{Jeong_criticalscaling}, In this situation the prefactor is not easily extracted and therefore we fit the expected algebraic behavior $J(1/T_1)_{+-}/\hbar=a (k_BT/J)^{-0.5}$ with a free fit parameter $a$. We obtain very good agreement of our numerical results with the fit using the value $a=0.71(2)$ in the entire regime of temperatures up to $k_BT/J\approx 2$, as shown in Fig.~\ref{fig:XXXtutto}. This means that our results are in agreement with the predictions of the quantum critical regime extending up to these temperatures. In addition in Fig.~\ref{fig:critmag} we offer a comparison between our magnetization data computed at finite temperature, and the scaling function close to (or on) a field-induced quantum critical point, which have been calculated and used in the literature\cite{Affleck_spin1_field,Sachdev_qaf_magfield,Jeong_criticalscaling}. At $h=2J=h_c$, from Eqs.~(2) and (3) in \onlinecite{Jeong_criticalscaling}, we know that the magnetization per site should behave as
\begin{align}
m(T)&=m_S-\left(\frac{2k_BT}{J}\right)^{d/2}\mathcal{M}(\delta h_c/k_BT)=\\
&=0.5-0.24312\sqrt{\frac{k_BT}{J}}
\end{align}
where $m_S=0.5$ is the magnetization per site at saturation, $d=1$ is the dimension of the system, $\delta h_c=0$ is the distance from the critical field and
\begin{equation}
\mathcal{M}(\delta h_c/k_BT)=\frac{1}{\pi}\int_0^{\infty}dx~\frac{1}{e^{x^2-(\delta h_c/k_BT)}+1}.
\end{equation}
We observe a good agreement between numerical results and the analytical prediction, which is supposed to be valid in the low temperature limit. We see MPS data approaching the analytics as the temperature is lowered.
In Fig.~\ref{fig:gappati} the results for the $1/T_1$ relaxation rate for the $S^{zz}$ correlations at a field of $h=5J$ are reported. The system at this magnetic field exhibits a gapped energy spectrum, and we denote the gap by $\Delta_g=h-2J=3J$. Due to the gapped energy spectrum an exponential decay with temperature is expected and indeed observed numerically.
In order to validate the exponential form for a different parameter regime we consider also the XX model at the same magnetic field which can be mapped onto free fermions. The corresponding gap in the energy spectrum is given by $\Delta_g=h-J=4J$. We computed the longitudinal part of the $1/T_1$ (the density-density correlation for the corresponding free fermions) analytically. Error bars come from the extrapolation, since correlations were evaluated up to a finite time. Also in this case a fit with an exponential function of $-\Delta_g$ is perfectly compatible with our calculations and validates our procedure.

\subsection{XXZ model}
\begin{figure}
  \centering
  \includegraphics[width=0.51\textwidth]{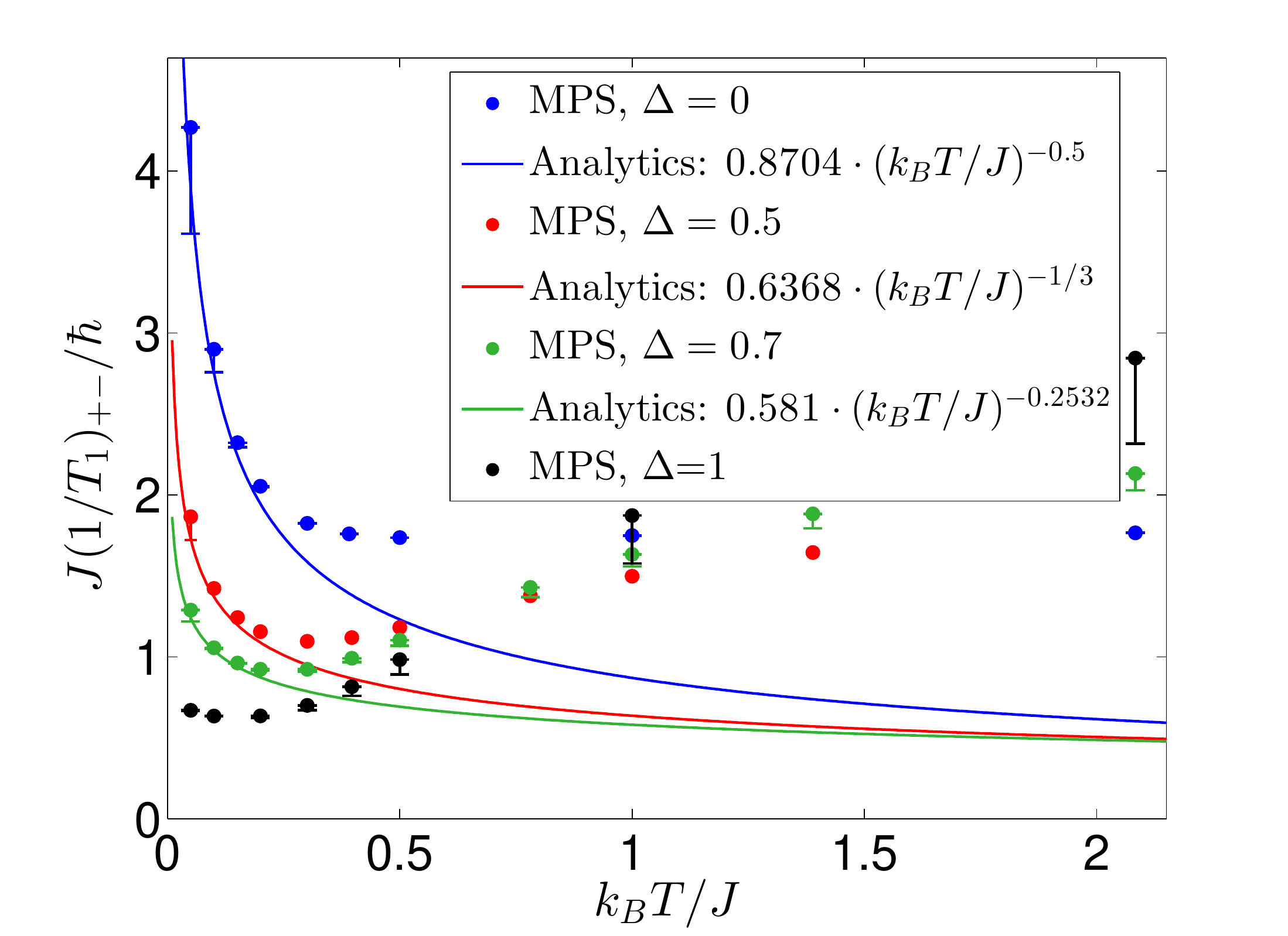}
  \caption{$(1/T_1)_{+-}$ as defined in Eq.~\ref{eq:nmrpm}, multiplied by $J/\hbar$ to have a dimensionless quantity, as a function of $k_BT/J$ for the XXZ model at different anisotropies. Dots with error bars are MPS results, solid lines are analytical predictions.}
  \label{fig:xxz}
\end{figure}

In order to explore more in detail the behavior of the relaxation time in the TLL phase, we move to the spin-1/2 XXZ model, Eq.~\ref{eq:XXZham}, in absence of magnetic field ($h=0$). For anisotropies $0\leq\Delta<1$ the ground state of this model is a TLL phase. As we discussed for the Heisenberg model at $h=J$, we expect that at low temperature the relaxation rate corresponding to the $S^{+-}$ correlations shows an algebraic divergence as given in Eq.~\ref{eq:oneovert1_analy}. We consider the cases $\Delta=0, \ 0.5, \ 0.7$ as shown in Fig.~\ref{fig:xxz}. The corresponding values of the Luttinger liquid parameters $K$ and $u$ are calculated using the Bethe ansatz formulas given e.g.~in Ref.~\onlinecite{giamarchi_book_1d}, while the amplitudes $A_x$ are taken from Ref.~\onlinecite{hikihara_amplitude_ladder}. Their rounded values are summarized in table \ref{tab:xxzpara}.
\begin{center}
\begin{table}
\begin{tabular}[t]{l||c|c|c}
$\Delta$ & $K$& $u$& $A_x$\\
\hline
0& 1&1&0.1471\\
0.5&0.75&1.299& 0.134\\
0.7& 0.6695& 1.4103& 0.1297\\
\end{tabular}
\caption{Values for the three relevant parameters $u$, $K$, and $A_x$ for different values of the anisotropy $\Delta$ in the XXZ model. $u$ has the units of $J/\hbar$ (lattice spacing equal to 1 here).}
\label{tab:xxzpara}
\end{table}
\end{center}
The agreement between the TLL prediction of the algebraic divergences and our numerical results is extremely good at low temperatures. For larger values of the anisotropy the divergence becomes weaker until for $\Delta=1$ one leaves the TLL region and no algebraic divergence is seen. As expected the predictions for the Luttinger liquid behavior disagree above a certain temperature of the order of $k_BT/J\approx 0.2$. Above this scale our numerical results show an upturn and the different curves even cross. Our results clearly show the importance for systems with small exchange constants to be able to go beyond the asymptotic expressions in order to make comparisons with the experiments.

\subsection{Dimerized model}
As a final example, we consider the isotropically dimerized spin-1/2 chain in presence of a magnetic field along the $z$ direction as defined in Eq.~\ref{eq:dimham}. This model can describe very well some interesting compounds like for example the copper nitrate $[\text{Cu}(\text{NO}_3)_2\cdot2.5\text{D}_2\text{O}]$, discussed in Refs.~\onlinecite{tennant03_dimerized,willenberg_dimerized}.
For this compound the coupling parameters are determined as  $J/k_B \approx 3.377$ K and $\delta J/k_B \approx 1.903$ K and we will focus on these strongly dimerized parameters in the following. The ground state of the system at $h=0$ has zero magnetization. A gap of $\Delta_g\sim 4.4~k_BK$  separates the ground state from the first excited state. In a magnetic field, the system shows a first quantum critical point at a magnetic field $h_{c1}$. At this point the system undergoes a transition from a gapped phase to a gapless, TLL phase. Here we focus on two cases: the gapped phase for $h=0$ and the TLL phase at $h\approx 1.01\cdot\Delta_g\gtrsim h_{c1}$.

In our numerical calculations we consider a chain of $L=130$ spins and in the imaginary time evolution a minimal truncation of $\varepsilon_{\beta}=10^{-20}$, a retained states maximum of 500, and a step of $\delta\beta=0.01474~J_w$. For the real time evolution we choose a minimal truncation of $\varepsilon_t=10^{-10}$, a time step amplitude $\delta t=0.0737~\hbar/J_w$ and a retained states maximum of 500 for temperatures $k_BT<0.68~J_w$, 800 for $0.68~J_w<k_BT<1.36~J_w$ and 2000 for higher temperatures. The maximal truncated weight is $10^{-6}$ in most cases, $10^{-5}$ for the highest temperatures. The $t_{\text{max}}$ reached still decreases from $59~\hbar/J_w$ for the lowest temperatures, to $15~\hbar/J_w$ for the highest ones, according to the requested precision. We calculate the relaxation time for the onsite correlation $S^{-+}$ (at $h=0$, $\langle S^{-+}\rangle=\langle S^{+-}\rangle=2\langle S^{zz}\rangle$).

\begin{figure}
  \centering
  \includegraphics[width=0.51\textwidth]{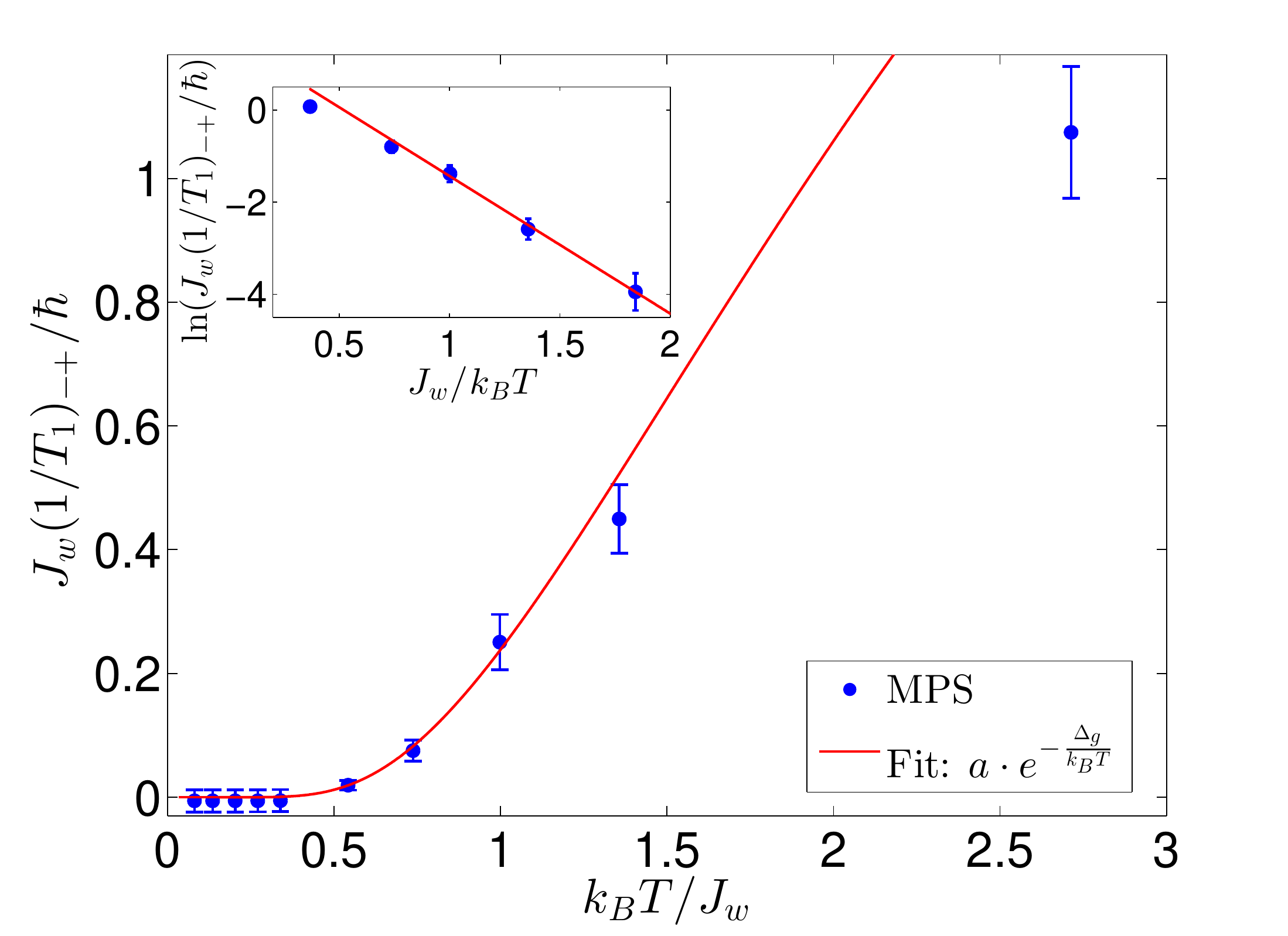}
  \caption{$(1/T_1)_{-+}$ as defined in Eq.~\ref{eq:nmrpm} multiplied by $J_w/\hbar$ to obtain a dimensionless quantity, plotted as a function of $k_BT/J_w$ for the isotropically dimerized spin-1/2 chain at $h=0$. Dots with error bars are MPS results, solid lines are combination of a fit with the analytical prediction. The fit parameter is $a\approx 4.7(3)$. In the inset, logarithmic representation of the same quantities when positive.}
  \label{fig:gap1}
\end{figure}

Due to the presence of a gap $\Delta_g$ in the absence of a magnetic field, the temperature dependence of the relaxation rate at low temperatures is expected to be exponentially activated. i.e.~$\propto e^{-\Delta_g/k_BT}$. In Fig.~\ref{fig:gap1} we show that our results agree very well with this exponential activation.

At larger temperatures $k_BT/J_w>1$ a saturation effect seems to set in. In the inset, lower temperature points have been cut because of the difficulties in the extrapolation which led to negative (though pretty close to 0) extrapolated values, as shown in the main panel of Fig.~\ref{fig:gap1}. A detailed description of the method used for the extrapolation and the association of an appropriate error bar is given in Appendix~\ref{app:errors}.

\begin{figure}
  \centering
  \includegraphics[width=0.51\textwidth]{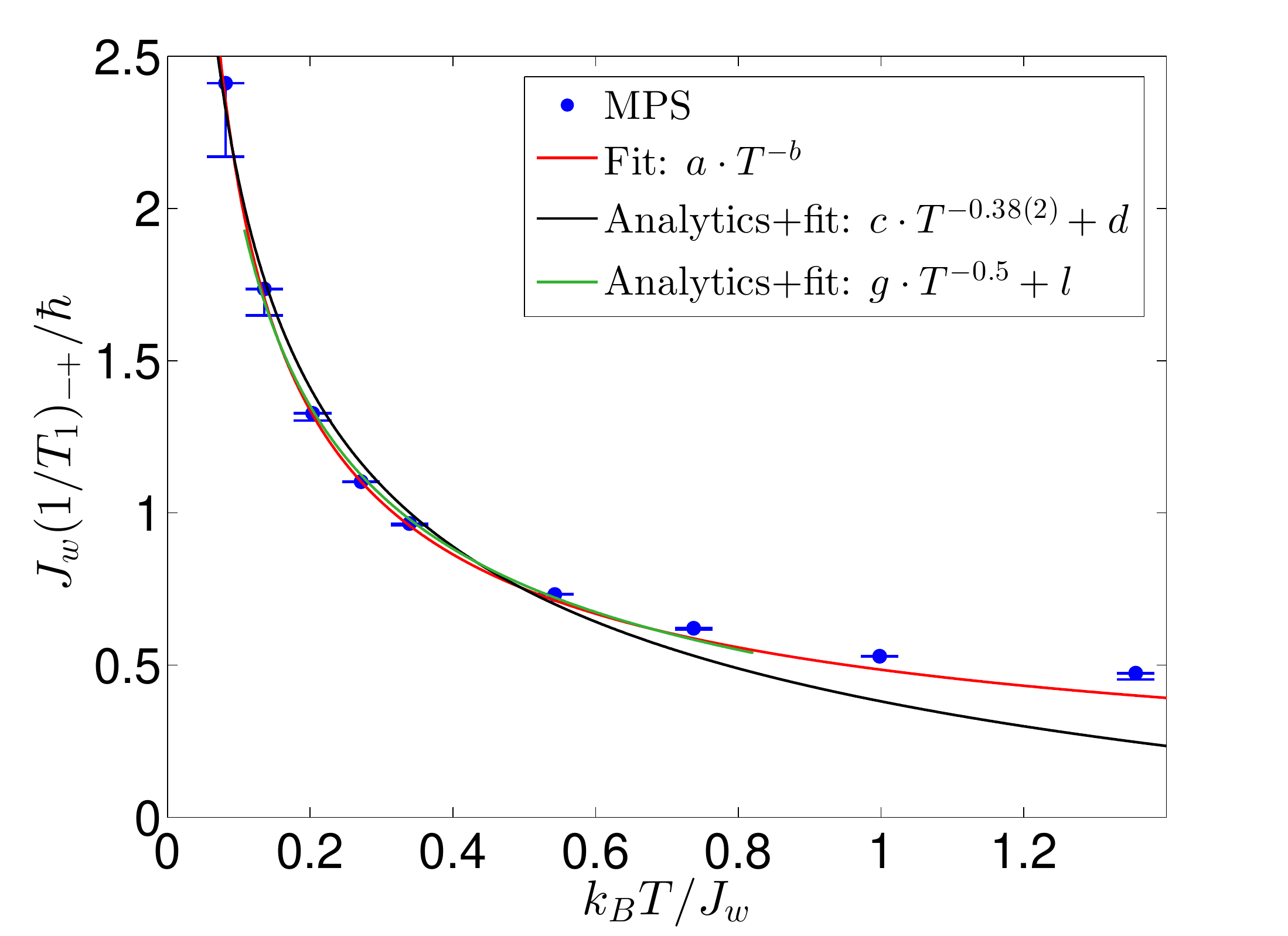}
  \caption{$(1/T_1)_{-+}$ as defined in Eq.~\ref{eq:nmrfinal} multiplied by $J_w/\hbar$ to obtain a dimensionless quantity, as a function of $k_BT/J_w$, for the isotropically dimerized spin-1/2 chain at $h\approx 3.02J_w$. Dots with error bars are MPS results, solid lines are fits or combinations of fit and analytics. The fit parameters are $a\approx 0.62(1)$, $b\approx 0.63(2)$, $c\approx 1.42(2)$, $d\approx -0.84(1)$, $g\approx 0.87(1)$, and $l\approx -0.25(1)$.}
  \label{fig:LLL}
\end{figure}

In contrast for the case $h\sim3.02J_w$ the low energy physics can be described by the TLL theory. Thus, the expected behavior of the relaxation rate as a function of the temperature is an algebraic divergence at low T of the form $\propto (k_BT/J_w)^{\frac{1}{2K} - 1}$. The TLL parameter $K$ which enters in this formula has been determined by separate calculations of the compressibility using MPS, and the flux dependence of the energy using infinite-size MPS calculations, giving $K\approx0.81(3)$. More details about this method can be found in Appendix \ref{app:LLpara}. The numerically obtained relaxation time is shown in Fig.~\ref{fig:LLL} and compared to the TLL predictions.

The black line represents the fit using the separately determined exponent $\frac{1}{2K} - 1$. A constant offset has been added since the behavior is not entirely dominated by the divergence. Deviations between the comparison of the analytical prediction and the numerical calculation are seen. We attribute these deviations to the proximity of the quantum critical point. In this regime, the TLL behavior is valid only for very low temperatures $k_B T\leq h-h_{c1}$. From our numerical results only the lowest temperature point lies within this region. To verify the influence of the quantum critical point, a fit using the critical power law shifted by a constant offset is performed. This fit leads for the intermediate temperature points to good results (see the green curve). A fit in which also the exponent is a fit parameter leads to an even larger exponent of $b\approx 0.63$ (red curve).

\section{Conclusions and outlook} \label{sec:conclusion}

Exploiting recent developments in finite temperature MPS techniques we computed the spin-lattice relaxation rate $1/T_1$ for a wide range of temperatures, for different Hamiltonians and for different quantum phases. In particular, we have considered the XX, Heisenberg, and XXZ Hamiltonians, plus the isotropically dimerized case. For the non-dimerized cases we have performed a detailed study of the gapless phase, the gapped phase and also of the quantum critical point. Numerical results were in very good agreement with analytical results available in the low-temperature limit. We have shown the deviation from the low-T law at finite temperature and we have swept through quantum critical points, situations in which theoretical results are more difficult to obtain. Our calculations prove that the MPS method can be successfully used to obtain the NMR relaxation time in regimes in which the field theoretical asymptotic values would not be applicable. The overlap between the regimes in which the numerical methods are applicable and the regime covered by the field theoretical asymptotic methods allows essentially a full description of the NMR behavior for the accessible regime of temperatures.

Having a method which can quantitatively compute the NMR relaxation time from a given microscopic Hamiltonian rather than simple asymptotic expressions should allow to test that the microscopic Hamiltonian does not miss an important term, and to fix the various coefficients by comparing the computed temperature dependence with the experimentally measured one. This is similar in spirit to what was achieved by the comparison of the computed neutron spectra with the measured ones for DIMPY\cite{schmidiger_dimpy_neutrons}. Another interesting direction is the investigation of the behavior of the relaxation mechanism of the spin excitations close to the quantum critical point. Indeed the nature of the relaxation mechanism is potentially different depending on whether one considers the $S^{zz}$ term or the $S^{\pm\mp}$ ones. For 3D systems a self energy analysis of the transverse part of $1/T_1$ was suggesting \cite{orignac07_bec} a behavior $1/T_1\propto e^{-3\Delta_g/k_B T}$ due to the necessity of making three magnon excitations to be able to scatter a magnon and get a finite lifetime while the $S^{zz}$ part leads, as shown in the present paper, to $1/T_1\propto e^{-\Delta_g/k_BT}$. Our numerical results which are able to correctly determine the exponential decay in the controlled cases of   the longitudinal excitations are thus potentially able to address this issue and potentially make contact on the experiments on that point \cite{mukhopadhyay_BPCB}. Such a study clearly going beyond the scope of the present paper, is thus left for future works.  

The present method works efficiently if the systems are one or quasi-one dimensional. One important challenge on the theoretical level is to extend the present analysis to the case of higher dimensional systems. In that case, although other methods such as quantum Monte-Carlo exist, the dynamical correlations in real time are still a challenge for which the MPS methods could bring useful contributions. Indeed the (numerically) rather complete knowledge of the one-dimensional correlation functions allow to incorporate them into approximation schemes such as RPA to capture a large part of the higher dimensional physics. Another route is to solve clusters of one dimensional structures, which allows to at least incorporate part of the transverse fluctuations. 

{\it Note added:} Just after submitting this work, a related numerical study by M.~Dupont, S.~Capponi and N.~Laflorencie appeared \cite{dupont16_nmr}. Our results are perfectly compatible with each other when comparison can be made. 

\begin{acknowledgements}
We acknowledge fruitful discussions with P. Bouillot and on NMR with C. Berthier and M. Horvati\'{c}. This work was supported in part by the Swiss NSF under Division II and the DFG and the ERC (Grant Number 648166, Phon(t)on).
\end{acknowledgements}

\appendix

\section{Extrapolation method}
\label{app:errors}
\begin{figure}
  \centering
  \includegraphics[width=0.51\textwidth]{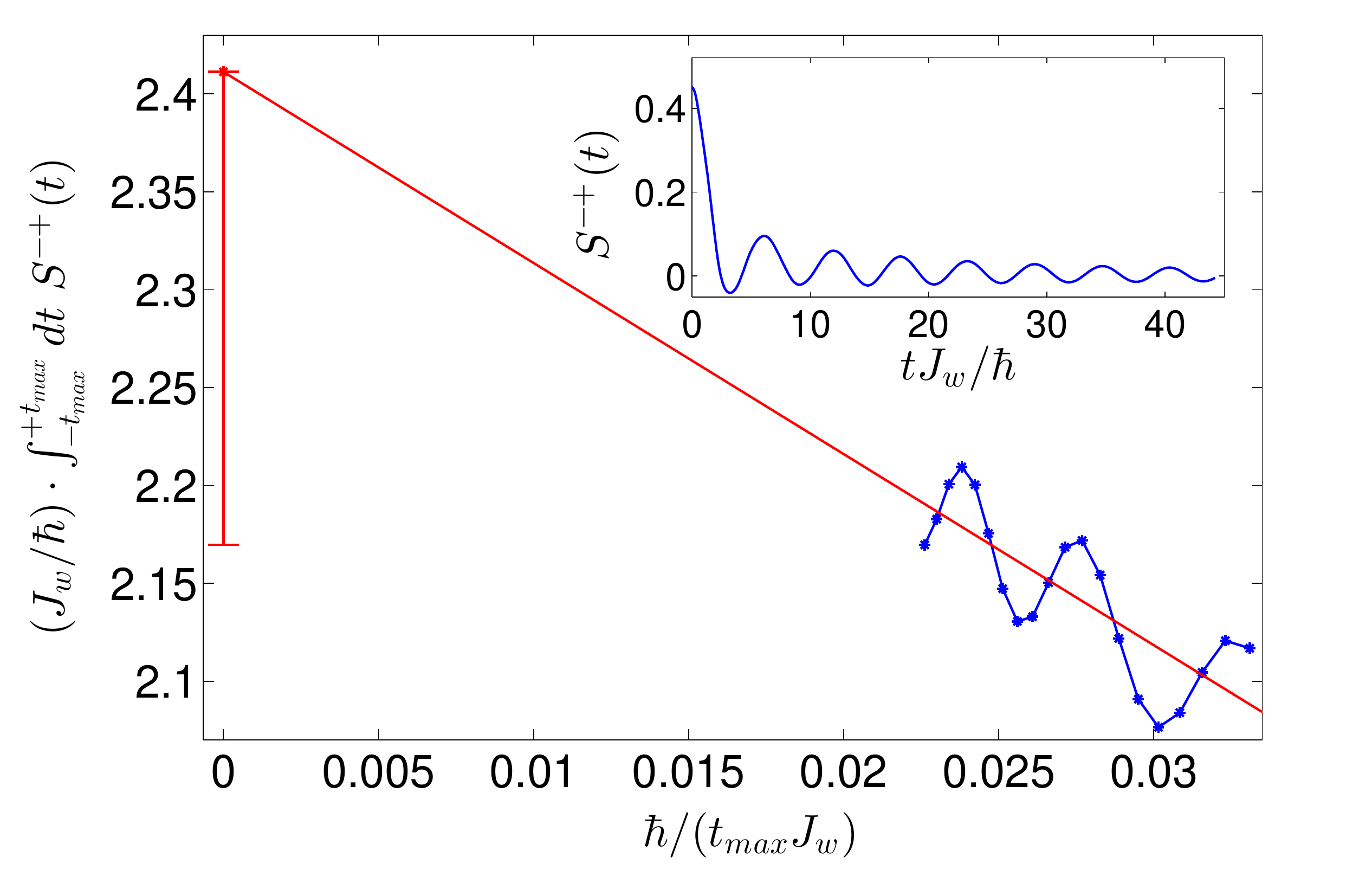}
  \caption{Integral over time from $-t_{\text{max}}$ to $+t_{\text{max}}$ of the onsite $S^{-+}$ correlations as a function of $\hbar/(t_{\text{max}}J_w)$ at $h\approx 3.02 J_w\gtrsim h_{c1}$, at the temperature $k_B T\approx 0.0814J_w$ for the dimerized model. The extrapolation is shown as a solid (red) line. The extrapolated point is reported with its error bar. The inset shows the correlations as a function of $tJ_w/\hbar$.}
  \label{fig:ll_extr}
\end{figure}

\begin{figure}
  \centering
  \includegraphics[width=0.51\textwidth]{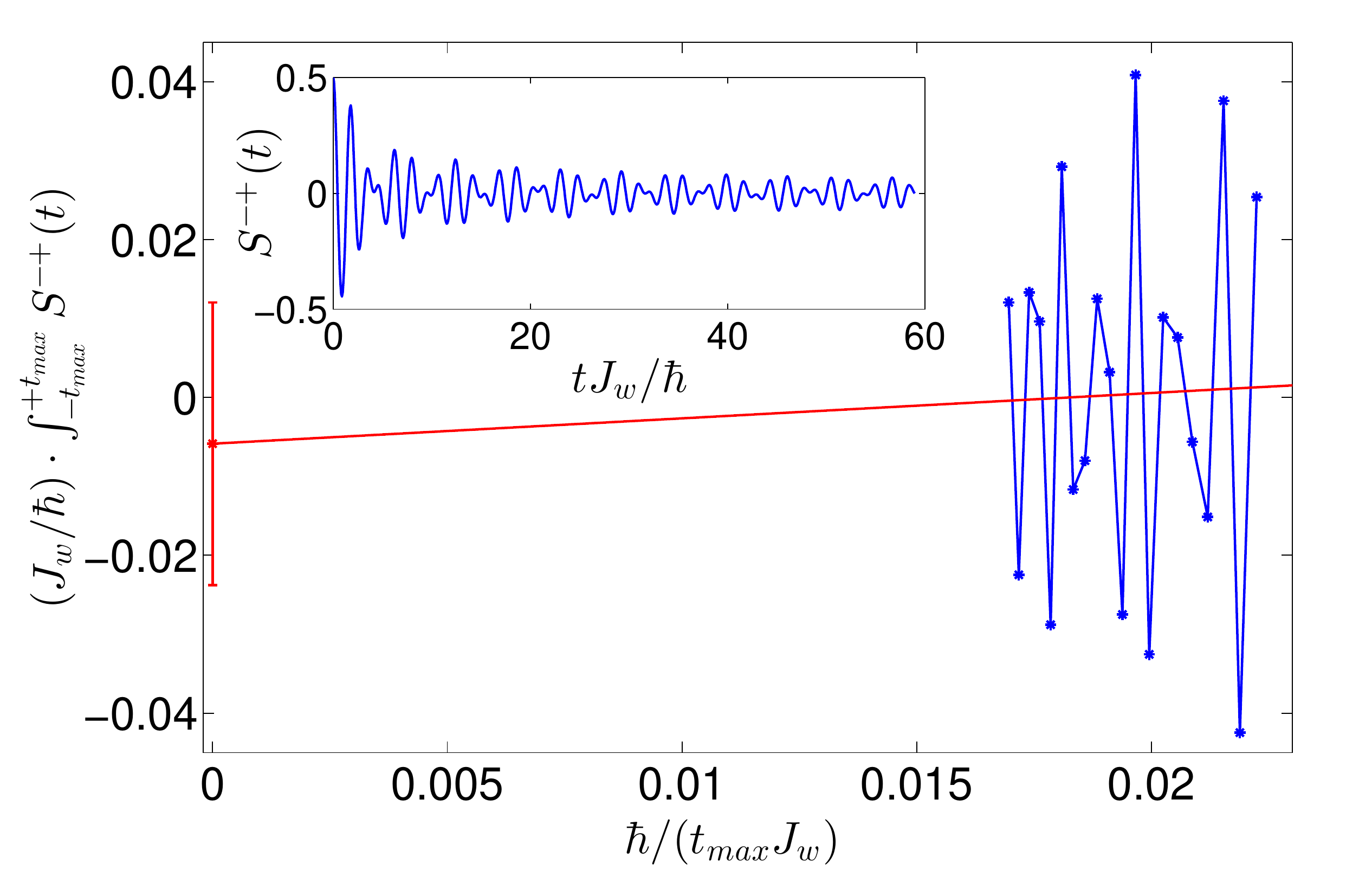}
  \caption{Integral over time from $-t_{\text{max}}$ to $+t_{\text{max}}$ of the onsite $S^{-+}$ correlations as a function of $\hbar/(t_{\text{max}}J_w)$ at $h=0$, at the temperature $k_B T\approx0.081J_w$ for the dimerized model. The extrapolation is shown as solid (red) line. The extrapolated point is reported with its error bar. The inset shows the correlations as a function of $tJ_w/\hbar$.}
  \label{fig:dim_extr}
\end{figure}

As discussed in Sec.~\ref{sec:procedure}, the numerical results for correlations are only available up to a certain time $t_{\text{max}}$. Since in principle the time integral of these correlations should be performed up to $\infty$, one needs to find a way to approximate the value of the extended integral and of the associated error bar. In order to do this we study the behavior of the integral as a function of $1/t_{\text{max}}$. We perform a linear fit of the value of the integral as a function of $1/t_{\text{max}}$ at the largest available values of $t_{\text{max}}$. We use this fit to extrapolate the value of the integral to $1/t_{\text{max}}\rightarrow 0$. If the value of the integral still shows a considerable trend, we associate to the extrapolated value a one-sided error bar corresponding to the difference between the extrapolated value and the value of the integral for the maximum $t_{\text{max}}$ available. An example is shown in Fig.~\ref{fig:ll_extr} for the dimerized chain in the TLL phase. For the case where the integral oscillates around a certain value and no clear trend is visible, we choose to associate a symmetric error bar with semi-amplitude equal to the distance between the extrapolated value itself and the value of the integral for the maximum $t_{\text{max}}$ available. An example is shown in Fig.~\ref{fig:dim_extr} for the dimerized chain in the gapped phase. In both cases, the extracted error bars should give a (most probably pessimistic) estimate of the uncertainty on the value of the integral.


\section{Consistency test using the XX model}
\label{app:XX}
To test the accuracy of the numerical procedure, we performed calculations for the XX model under a magnetic field along the $z$ direction:
\begin{equation}
H=\frac{J}{2}\sum_j\left(S^+_jS^-_{j+1}+\text{h.c.}\right)-h\sum_jS^z_j.
\label{eq:modelxx}
\end{equation}
 For this specific model we focus on two specific cases, $h=0$ (gapless phase) and $h=5J$ (gapped phase), and on $S^{zz}$ correlations. In particular, we determine for different temperatures the ratio $1/T_1$ for $S^{zz}$ correlations, which we define here as:
\begin{equation}
\left(\frac{1}{T_1}\right)_{zz}=\int\limits_0^{t_{\text{max}}}dt\quad \operatorname{Re}S^{zz}_{j,T}(t).
\label{eq:testnmr}
\end{equation}
We compare our numerical results obtained via the described procedure using MPS, with exact analytically results \cite{giamarchi_book_1d}. In the limit of an infinite-size system, the exact result for the onsite correlations at a temperature $T$ and $h=0$ is given by
\begin{multline}
S^{zz}_{j,T}(t)=\frac{J_0\left(Jt/\hbar\right)}{2\pi}\cdot\int\limits_{-\pi}^{+\pi}dk \ e^{i\lambda_kt/\hbar}\cdot f_k(\beta) \ -\\
-\frac{1}{4\pi^2}\cdot\left| \ \int\limits_{-\pi}^{+\pi}dk \ e^{i\lambda_kt/\hbar}\cdot f_k(\beta) \ \right|^2,
\label{eq:h0}
\end{multline}
where $J_0(\dots)$ is the 0th-order Bessel function of the first kind, $i$ is the imaginary unit, $\lambda_k=J\cos{(k)}$, where $k$ is the dimensionless momentum and
\begin{equation}
f_k(\beta)=\frac{1}{1+e^{\beta\lambda_k}}
\end{equation}
is the Fermi function, where $\beta$ is the inverse temperature.

For $h\neq0$ one obtains
\begin{multline}
S^{zz}_{j,T}(t)=\frac{J_0\left(Jt/\hbar\right)}{2\pi}\cdot e^{iht/\hbar}\cdot\int\limits_{-\pi}^{+\pi}dk \ e^{i\lambda'_kt/\hbar}f_k(\beta) \ -\\
-\frac{1}{4\pi^2}\cdot\left| \ \int\limits_{-\pi}^{+\pi}dk \ e^{i\lambda'_kt/\hbar}f_k(\beta) \ \right|^2+~\frac{1}{4}~+\\
+\frac{1}{4\pi^2}\cdot\left[\int\limits_{-\pi}^{+\pi}dk \ f_k(\beta)-2\pi\right]\cdot\int\limits_{-\pi}^{+\pi}dk \ f_k(\beta).
\label{eq:h5}
\end{multline}
Here $\lambda'_k=J\cos{(k)}-h$ and $f_k(\beta)=\frac{1}{1+e^{\beta\lambda'_{k}}}$.

\begin{figure}
  \centering
  \includegraphics[width=0.52\textwidth]{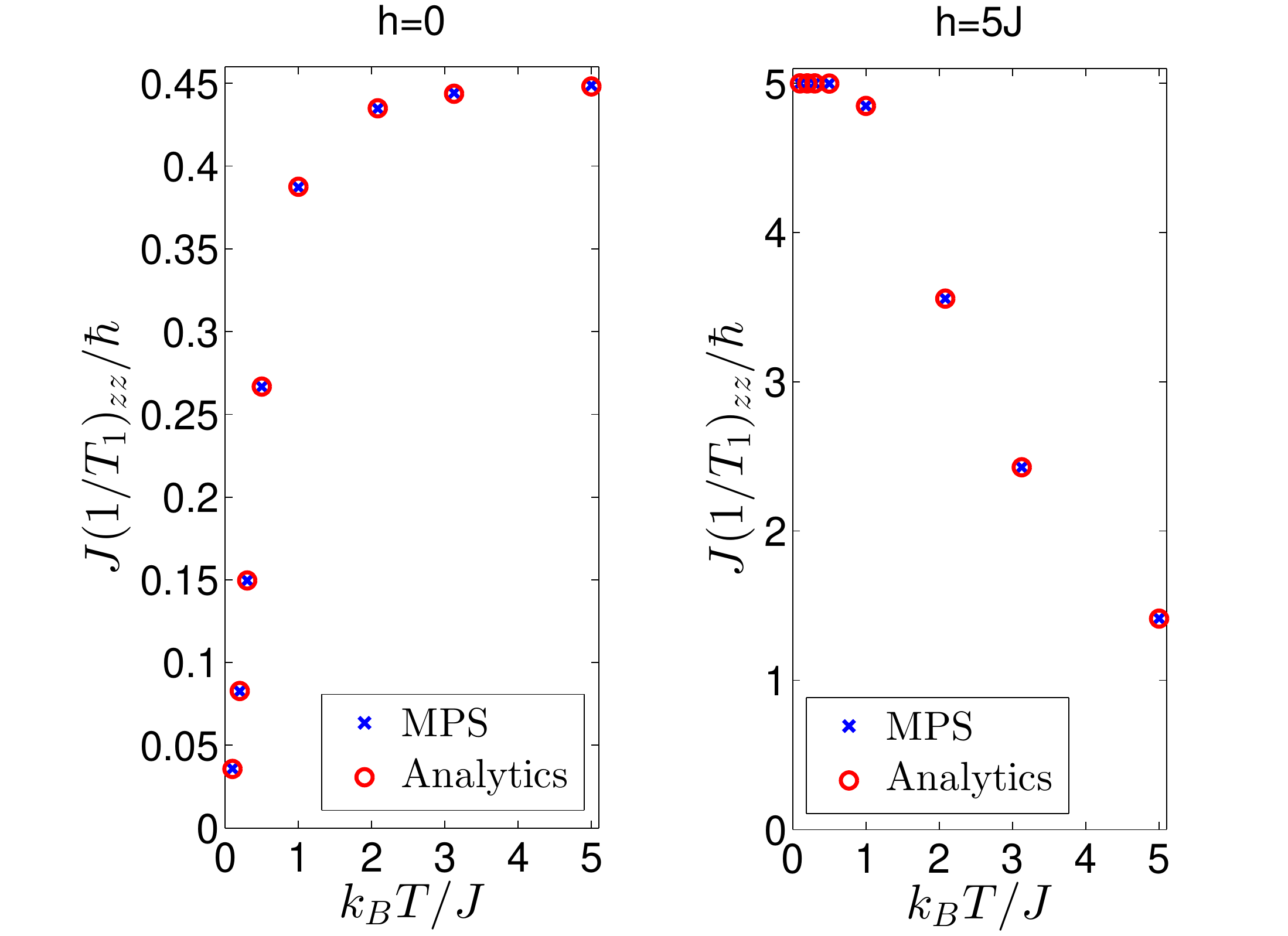}
  \caption{$(1/T_1)_{zz}$ as defined in Eq.\ref{eq:testnmr} (multiplied by $J/\hbar$) as a function of $k_BT/J$ for the XX model under a magnetic field $h=0$ and $h=5J$. The analytical results correspond to infinite system size and $t_{\text{max}}=20\hbar/J$. The numerical results are obtained with $t_{\text{max}}=20\hbar/J$ and $L=100$. The agreement found is excellent.}
  \label{fig:xx}
\end{figure}

As for the numerical procedure, simulations are performed for a chain of $L=100$ spins. Onsite correlations are measured in the center of the chain ($j=50$) to avoid boundary effects. Imaginary time evolutions are performed using the following parameter set: minimal truncation $\varepsilon_\beta=10^{-20}$, retained states maximum 400 and step $\delta\beta=0.01~J^{-1}$. Real time evolutions are performed using: minimal truncation $\varepsilon_t=10^{-10}$, a retained states maximum of 800, maximal truncated weight $10^{-6}$, and time step $\delta t=0.05 \hbar/J$ up to $t_{\text{max}}=20\hbar/J$. Results of the comparison theory-numerics are reported in Fig.~\ref{fig:xx}. The agreement between the analytical and the numerical results is extremely good at all temperatures, which justifies our procedure.

\section{Determination of the TLL parameters}
\label{app:LLpara}
To get the values of the TLL parameters $u$ and $K$ we determine first their ratio $K/u$, which is related to the static TLL susceptibility, and their product $u\cdot K$, related to the variation of the energy with a flux. Then, the two values of $u$ and $K$ trivially follow by recombination of the two previous results.

The ratio $K/u$ is determined from the static TLL susceptibility of the system according to the relation\cite{giamarchi_book_1d}
\begin{equation}
\frac{K}{u}=\frac{\pi}{L\frac{d^2E_0}{dM^2}},
\end{equation}
where $L$ is the size of the system, $E_0$ is the ground state energy, and $M$ is the total magnetization.  The second derivative has to be discretized since $M$ in a spin-1/2 system can only vary by integer steps (thus $\Delta M=1$):
\begin{equation}
\frac{K}{u}(M)=\frac{\pi}{L\left[E_0(M+1)+E_0(M-1)-2E_0(M)\right]}.
\end{equation}
$E_0$ can be evaluated at fixed values of magnetization via standard finite-size DMRG. The magnetization, defined at the beginning of the simulation by the initial distribution of spins, is set as a conserved quantum number.

The product $u\cdot K$ can be determined by studying the variation of the ground state energy of the system in response to a variation of a flux through the system.
To be more precise, for a fixed value M of the magnetization\cite{giamarchi_book_1d}
\begin{equation}
uK(M)=\pi L\frac{d^2E_0(\Phi,M)}{d\Phi^2}\bigg |_{\Phi=0},
\label{eq:utimesk}
\end{equation}
The flux is represented by twisted periodic boundary conditions, $\Psi(L)=\Psi(0)\cdot e^{i\Phi}$. This condition can be transferred into the Hamiltonian via the following transformation:
\begin{align}
S^+_jS^-_{j+1} &\longrightarrow S^+_jS^-_{j+1}\cdot e^{i\frac{\Phi}{L}},\nonumber\\
S^-_jS^+_{j+1} &\longrightarrow S^-_jS^+_{j+1}\cdot e^{-i\frac{\Phi}{L}},
\end{align}
which distributes homogeneously the total flux $\Phi$ along the chain. For each fixed value of $M$ (and therefore of $h$), we evaluate $E_0(\Phi)\vert_M$ within an infinite-size MPS algorithm for symmetric values of $\Phi$ around zero. The resulting ground state energy $E_0(\Phi)$ close to $\Phi=0$ can be approximated by a parabola and we fit the points with a second degree polynomial of the form $P(\Phi)=a_M\Phi^2+b_M\Phi+c_M$, where $a_M, b_M$ and $c_M$ are the fit parameters. According to Eq.~\ref{eq:utimesk}, the fitting parameter $a_M$ is related to the product
\begin{equation}
uK(M)=2\pi La_M.
\end{equation}

\end{document}